\documentstyle[aps,12pt]{revtex}
\draft
\title{Spatial evolution of short laser pulses under coherent
population trapping}
\author{V.G.Arkhipkin}
\address{Institute of Physics, Russian Academy of Sciences, 660036
Krasnoyarsk ; Krasnoyarsk State Technical University, 660074
Krasnoyarsk, Russia}
\author{I.V.Timofeev}
\address{Krasnoyarsk state university,660041 Krasnoyarsk, pr. Svobodny, 79}
\begin{document}
\maketitle

\begin{abstract}

Spatial and temporal evolution is studied of two powerful short
laser pulses having different wavelengths and interacting with a
dense three-level Lambda-type optical medium under coherent
population trapping. A general case of unequal oscillator
strengths of the transitions is considered. Durations of the probe
pulse and the coupling pulse $T_{1,2}$ ($T_2>T_1$) are assumed to
be shorter than any of the relevant atomic relaxation times. We
propose analytical and numerical solutions of a self-consistent
set of coupled Schr\"{o}dinger equations and reduced wave
equations in the adiabatic limit with the account of the first
non-adiabatic correction. The adiabaticity criterion is also
discussed with the account of the pulse propagation. The dynamics
of propagation is found to be strongly dependent on the ratio of
the transition oscillator strengths. It is shown that envelopes of
the pulses slightly change throughout the medium length at the
initial stage of propagation. This distance can be large compared
to the one-photon resonant absorption length. Eventually, the
probe pulse is completely reemitted into the coupling pulse during
propagation. The effect of localization of the atomic coherence
has been observed similar to the one predicted by Fleischhauer and
Lukin (PRL, {\bf 84}, 5094 (2000).

\end{abstract}


\section{INTRODUCTION}

The electromagnetically induced transparency (EIT) can be used to
make optically thick media transparent to resonant laser radiation
\cite{Harr}. The EIT is the result of various quantum interference
effects such as nonlinear interference \cite{Raut}, coherent
population trapping (CPT) \cite{Agap,Arim}, adiabatic population
transfer (APT) \cite{Berg}. Optical characteristics of the matter
undergo drastic changes under those effects to such an extent that
they can now be manipulated. A lot of interesting applications
based on that have been proposed and experimentally realized (see,
e.g. \cite{Agap,Arim,Berg,Yam,Hau,Kash,Phillips,Liu}).

Interesting and unusual phenomena caused by the above indicated
effects can be observed when laser pulses propagate in a resonant
three-level medium. Propagation of pulses under EIT conditions was
studied for example in \cite{1,2,3,4,5,6,7,7a}. As a rule, the
situations are considered when both pulses have identical forms
and their duration is longer than the relaxation time of the
intermediate resonant state (matched pulses \cite{1}; dressed
field pulses \cite{2,3}) or when the duration of the coupling
radiation considerably exceeds that of the probe radiation
(adiabatons \cite{5,6,Kasapi}). A theoretical study of certain
features of spatial evolution under APT conditions is presented in
\cite{6,7,8}.  Propagation of soliton-like pulses in a three-level
system is studied in \cite{Hioe}. A three-level system with equal
oscillator strengths is under consideration in all the above
mentioned studies, whereas in actual fact the transition
oscillator strengths most often are different.

In this paper, the spatial and temporal evolution is studied of
two overlapping short laser pulses propagating in a resonant
optically thick medium that consists of three-level
$\Lambda$-atoms. Pulses of such configuration are widely used to
enhance the efficiency of nonlinear generation processes
\cite{9a,9b}. The two pulses are assumed to have identical shapes
but different durations $(T_2>T_1)$ as shown in Fig.1. The pulse
durations are much shorter than any of the times of relaxation in
the medium (short pulses). It is also assumed that the pulse
envelopes satisfy the adiabaticity criterion \cite{3,10,11}:

\begin {eqnarray} \label {e1}
(G_2\dot{G_1}-G_1\dot{G_2})/G^3\ll1,
\end {eqnarray}
where $G_{1,2}$ are the  Rabi frequencies of the respective
fields, $G=\sqrt{G_1^2+G_2^2}$; the dot refers to time
derivatives. Condition (\ref{e1}) is easy to satisfy by making one
of the pulse amplitudes or both of them large even for short
pulses. This will induce a strong coherence at the Raman
transition resulting in the effect of CPT. The latter considerably
decreases the absorption of the propagating resonant pulses. The
dynamics of propagation of such pulses is studied here without
restriction of the relationship between oscillator strengths of
transitions.

Our theoretical model involves a set of coupled Schr\"{o}dinger
equations and a set of reduced wave equations, allowing thus a
simultaneous description of temporal and spatial evolution of the
atomic system and the radiation. The equations are analyzed in
approximation (\ref {e1})  with the account of the first
non-adiabatic correction. It will be shown that the dynamics of
propagation strongly depends on the oscillator strengths ratio.
Also the spatial and temporal behavior is analyzed of the atomic
Raman coherence. A possibility of localizing the atomic coherence
spatially has been established. The results obtained are compared
to the results reported in \cite{Lukin}.

The paper is organized as follows. In Section II, we describe the
model and present the basic equations. Section III contains
self-consistent solutions of those equations in the adiabatic
limit and describes the temporal behaviour of the level
populations and the atomic Raman coherence in the optically thin
medium. The spatial evolution of pulses in the optically thick
medium is described in Section IV for various oscillator strength
ratios. In Section IV, we also discuss the adiabaticity criterion
and demonstrate the effect of spatial localization of the atomic
coherence. Finally we summarize the results obtained.

\section{BASIC EQUATIONS}

The three-level system under consideration is shown in Fig.1
together with the temporal configuration of the pulses as they
enter the medium. The pulses travel along the same direction, $z$.
States $|0\rangle$, $|1\rangle$ and $|2\rangle$ are connected by
laser pulses $E_1=1/2\,E_1(t)\ exp[-i(\omega_1 t-k_1 z)]+c.c.$ and
$E_2=1/2\,E_2(t)\ exp[-i(\omega_2 t-k_2 z)]+c.c.$, respectively.
In our further consideration we shall refer to the first pulse
$E_1$ as the probe and to the second pulse $E_2$ as the coupling
pulse. The Rabi frequency of probe pulse is comparable
with that of the coupling pulse. The pulses are sent simultaneously
into an atom. The pulse durations $T_{1,2}$ $(T_2>T_1)$ are
assumed to be much less than any of the relaxation times of atoms.
The transition $|0\rangle-|2\rangle$ is electric dipole forbidden.
The intermediate state $|1\rangle$  is in a one-photon resonance
with each field interacting only with the respective transition.

The following standard set of equations describes the spatial and
temporal dynamics of the probability amplitudes of atomic states
$b_{0,1,2}$ and slowly varying Rabi frequencies
$G_1=d_{10}E_1(t)/2\hbar$, $G_2=d_{21}E_2(t)/2\hbar$ in the
local-time coordinate system $\tau = t-z/c$:

\begin {eqnarray} \label {eq1}
\frac {\partial b_0}{\partial \tau}& =& iG_1^*b_1\exp {(-ik_1 z)},\nonumber \\
\frac {\partial b _ 2} {\partial \tau}& =& iG_2^{*} b_1\exp {(-ik_2z)}, \nonumber \\
\frac {\partial b _ 1} {\partial \tau}& =& iG_1b_0\exp {(ik_1 z)} +
iG_2b_2\exp {(ik_2 z)}.
\end {eqnarray}

\begin {eqnarray} \label {eq2}
\frac {\partial G_{1}} {\partial z}& =& iK_1 b_1b_0^*\exp {(ik_1 z)},\nonumber \\
\frac {\partial G_{2}} {\partial z}& =& iK_2 b_1b_2^*\exp{(ik_2z)}.
\end {eqnarray}

Here we assumed zero one-photon detunings. $K_{1,2} =
{\pi\omega_1\left|d_{10,12}\right|^2}N/\hbar c$ are the
propagation coefficients, $N$ is the atomic concentration,
$d_{10,12}$ are the dipole transition matrix elements,
$\omega_{1,2}$, $k_{1,2}$ are the frequencies and the wave numbers
of the interacting waves in vacuum, $c$ is the light velocity in
vacuum. All atoms are assumed to be initially in the ground state
$|0\rangle$: $b_0(-\infty,z)=1$, $b_{1,2}(-\infty,z)=0$. We use
Gaussian pulses at the medium entrance $z=0$ for the purpose of
numerical simulation: $G_1(\tau)=G_1^0 \exp(-\tau^2ln2/T_1^2)$,
$G_2(\tau)=G_2^0 \exp[-\tau^2ln2/T_2^2]$.

In terms of $a_0=b_0\exp{(i k_1 z)}$, $a_2=b_2\exp{(i k_2 z)}$,
$a_1=i b_1$, equations ({\ref{eq1}}) and ({\ref{eq2}}) can be
written as:

\begin {eqnarray} \label {equ1}
\frac {\partial a_0} {\partial\tau}& =& G_1^{*} a_1, \nonumber \\
\frac {\partial a_2}{\partial\tau}& =& G_2^{*}a_1, \nonumber \\
\quad\frac{\partial a_1}{\partial\tau}& =& - G_1 a_0 - G_2 a_2.
\end {eqnarray}
\begin {eqnarray} \label {equ2}
\frac {\partial G_{1}} {\partial z}& =&  -K_1 a_1 a_0^{*},\nonumber \\
\frac {\partial G_{2}} {\partial z}& =&  -K_2 a_1 a_2^{*}.
\end {eqnarray}

The coupled equations ({\ref{equ1}}) and ({\ref{equ2}}) give a
complete semiclassical description of the resonant
different-wavelengths propagation problem we are dealing with.

\section{TEMPORAL DYNAMICS OF LEVEL POPULATIONS AND RAMAN COHERENCE
IN THE ADIABATIC APPROXIMATION (OPTICALLY THIN MEDIUM)}

In this section we study the temporal dynamics of populations and
the atomic coherence in the given time-dependent field, assuming
that the medium is optically thin. Based on that, $G_{1,2}$ will
not depend on the coordinate $z$. One can show that condition
(\ref{e1}) for Gaussian pulses reduces to $G_2^0 T_1\gg1$ when
$T_2/T_1>\sqrt{2}$. With the first non-adiabatic correction, the
solution of Eq. ({\ref{equ1}}) takes on the form:

\begin {eqnarray} \label {eq3}
a_0 &\simeq& \frac{G_2(\tau)}{G(\tau)}, \quad
a_2\simeq-\frac{G_1(\tau)}{G(\tau)}, \nonumber \\
a_1 &\simeq& \frac {1}{G_1}\frac{\partial(G_2/G)}
{\partial\tau}\simeq
-\frac{1}{G_2}\frac{\partial(G_1/G)}{\partial\tau},
\end {eqnarray}
where $G(\tau)=\sqrt{G_1^2(\tau)+G_2^2(\tau)}$.

The solutions for probability amplitudes are convenient to be
represented as:

\begin {eqnarray} \label {eq3a}
a_0=\cos{\theta(\tau)}, \quad a_2=-\sin{\theta(\tau)},
\end {eqnarray}
where the mixing angle $\theta(\tau)$ is defined as
$\tan{\theta(\tau)}=G_1(\tau)/G_2(\tau)$ (we shall discuss its
meaning later on).

The expression for $a_1$ can be reduced to:

\begin {eqnarray} \label {eq4}
a_1=(G_2\dot{G_1}-G_1\dot{G_2})/G^3=\dot{\theta}/G.
\end {eqnarray}
In adiabatic limit (\ref {e1}) $|a_1|=|\dot{\theta}/G|\ll1$
($\dot{\theta}=\partial{\theta}/\partial{\tau}$), i.e. the
population of the intermediate state $|1\rangle$ is close to zero
all the time during the interaction with pulses. This also implies
that the resonant absorption of the light pulses is weak
(electromagnetically induced transparency) and the population is
mainly distributed between the initial $|0\rangle$ and the final
$|2\rangle$ states:

\begin {eqnarray} \label {eq5}
|a_0|^2+|a_2|^2\simeq 1.
\end {eqnarray}
Equality  (\ref {eq5}) reflects the fact that atoms are trapped in
the CPT state:
$a_{cpt}=(G_2/G)a_0-(G_1/G)a_2=a_0\cos{\theta}-a_2\sin{\theta}=1$.
This effect is responsible for the decrease in the resonant
absorption of propagating pulses. Also the Raman coherence
$\rho_{20}=a_0a_2^*$ occurs:

\begin {eqnarray} \label {eq5a}
\rho_{20}=-\frac{1}{2}\sin{(2\theta)}  \quad \mbox{or}  \quad
\rho_{20}=-\frac{G_1G_2}{G_1^2+G_2^2}.
\end {eqnarray}

Obviously, the maximum coherence (in absolute value)
$|\rho_{20}|=1/2$ is reached when $\theta=\pi/4$ ($G_1^0=G_2^0$).
Figure 2 shows the temporal behavior of level populations
$|a_{0,2}(\tau)|^2$, the atomic Raman coherence
$|\rho_{20}(\tau)|$ and the mixing angle $\theta(\tau)$ for
Gaussian pulses in the optically thin medium.

The above results can be interpreted in terms of the
three-dimensional vector model where vector variables
$\vec{a}=(a_0,a_2,a_1)$ and $\vec{G}=(G_2,G_1,0)$ (the torque
vector) are introduced. Using these variables, we can rewrite (\ref
{equ1}) as

\begin {eqnarray} \label {eq7}
\dot {\vec {a}} = \vec {G} \times\vec {a},
\end {eqnarray}
where the sign '$\times$' means the vector product. The solution
of equation (\ref {eq7}) is vector $\vec{a}=(G_2/G,-G_1/G,
\dot{\theta}/G)$. Components of the vector $\vec{a}$ coincide with
adiabatic solution (\ref {eq3}). Figure 3 illustrates the dynamics
of vectors $\vec{G}$ and $\vec{a}$ in the three-dimensional vector
model. The torque vector $\vec{G}$ moves in the
$\vec{e}_1$-$\vec{e}_2$ plane. And vector $\vec{a}$ having a small
angle with respect to vector $\vec{G}$ ($|\dot{\theta}/G| \ll 1$)
follows it. Thus one can see an absolute analogy with the adiabatic
following in the case of a light pulse interacting with a
two-level atom \cite{13}. Such a simple picture can be observed
only in optically thin media. In optically dense media, $G_{1,2}$
and hence $\theta$ become dependent on the $z$ coordinate.

\section{SPATIAL EVOLUTION OF INTERACTING PULSES IN OPTICALLY DENSE MEDIA}

\subsection{General case:
 unequal oscillator strengths $(K_1\neq K_2)$}

The condition $K_1\neq K_2$ means that the probability of the
$|0\rangle-|1\rangle$ transition is not equal to that of the
$|2\rangle-|1\rangle$ transition. We note that in the ideal
adiabatic limit $a_1=0$, and the pulses would not change their
shape as they propagate in a medium that is optically thick for
each of the pulses (see (\ref {equ2})). However this is not the
case. The non-adiabatic correction has to be introduced for real
situations, which results in the induced dipole moments at the
transitions $|1\rangle-|0\rangle$ and $|1\rangle-|2\rangle$ , and
in the change of both pulses traveling in the medium. In order to
attribute this effect to propagation of the interacting pulses in
an optically dense medium, it is necessary to solve Eqs. (\ref
{equ1}) and (\ref {equ2}) in a self-consistent way.

Use (\ref {eq3a}) and (\ref {eq4}) to rewrite field equations
(\ref {equ2}) in the following form

\begin {eqnarray} \label {eq8}
 \frac{\partial G_1} {\partial z} =
  -K_1\frac{\dot{\theta}}{G}\cos{\theta}, \qquad
 \frac{\partial G_2}{\partial z} =
  K_2 \frac{\dot{\theta}}{G}\sin{\theta}.
\end {eqnarray}

From (\ref {eq8}), one can show that

\begin {eqnarray} \label {eq8a}
K_2 G_1^2(\tau,z) &+& K_1 G_2^2(\tau, z) \nonumber \\
                  &=& K_2 G_1^2(\tau,z=0)+K_1G_2^2(\tau,z=0),
\end {eqnarray}
i.e. $K_2 G_1^2(\tau,z)+K_1 G_2^2(\tau, z)$ does not depend on the
$z$ coordinate.

Equation  (\ref {eq8a})  describes the Manley-Raw relation, i.e.
the law of conservation of the total energy density during
propagation under CPT conditions.

Using the definitions of $\theta$ and $G$, we obtain the following
expressions for $G_{1,2}(\tau,z)$

\begin {eqnarray} \label {eq3b}
G_1(\tau,z)=G(\tau,z)\sin{\theta(\tau,z)}, \nonumber \\
G_2(\tau,z)=G(\tau,z)\cos {\theta(\tau,z)}.
\end {eqnarray}

Substitution (\ref {eq3b}) into the Manley-Raw relation gives the
following expression for $G(\tau,z)$:

\begin {eqnarray} \label {eq10}
G^2(\tau,z)=G^2_0(\tau)\frac{K_2\sin^2(\theta_0(\tau))+
K_1\cos^2(\theta_0(\tau))}{K_2\sin^2(\theta(\tau,z))+
K_1\cos^2(\theta(\tau,z))},
\end {eqnarray}
where $G_0^2(\tau)=G_{1}^2(\tau,0)+G_{2}^2(\tau,0)$ and
$\theta_0(\tau)=\theta_0(\tau,0)$ are the functions at the medium
entrance $z=0$.

So the dynamics of level populations as well as the atomic Raman
coherence and the evolution of the pulse shape are completely
determined by the function $\theta(\tau,z)$ that depends on both
the time $\tau$ and the coordinate $z$. Differentiating
$\tan{\theta}=G_1/G_2$ on the $z$ and using (\ref {eq8}) we obtain
the following equation for $\theta(\tau,z)$:

\begin {eqnarray} \label {eq11}
\frac{\partial\theta}{\partial\tau}+\frac{G^2(\tau,z)}{K(\theta)}
\frac{\partial\theta}{\partial z}=0,
\end {eqnarray}
where $K(\theta(\tau,z))=K_1\cos^2{(\theta(\tau,z))}+
K_2\sin^2{(\theta(\tau,z))}$.

Eq.(\ref {eq11}) is similar to the equations describing nonlinear
waves with the sharpening of the wave front during propagation
\cite{14}. The parameter $u=G^2/K$ can be treated as the
"nonlinear" velocity. The nonlinear velocity can be described as
$u(\tau,z)=A(\tau)/K^2(\theta(\tau,z))$ where the first factor
$A(\tau)=
G_0^2(\tau)[K_2\sin^2{(\theta_0(\tau))}+K_1\cos^2{(\theta_0(\tau))}]$
is independent of the $z$ coordinate, and the second factor
$K(\theta(\tau,z))$ is maintained along a characteristic of
Eq.(\ref {eq11}) $\theta(\tau,z)=const$. This allows to write down
the characteristic curve equation in an evident form:

\begin{eqnarray} \label {eq12}
z=\frac{1}{K^2(\theta_0)}\int_{\tau_0}^{\tau}A(\tau')d\tau'
\end{eqnarray}
Here $\tau_0$ is the time at which the characteristic curve goes
out of the medium boundary.

The solution for $\theta(\tau,z)$ has the form:

\begin{eqnarray} \label {eq13}
\theta(\tau,z)=\theta_0(\tau_0,0)
\end{eqnarray}
Here $\tau_0$  is to be determined from Eq.(\ref {eq12}).

\subsection{The case of equal oscillator strengths $(K_1=K_2)$}

In the case of equal oscillator strengths $K_1=K_2\equiv K$ the
$G(z,\tau)$ function is not subject to changes during propagation
($G(\tau,z)=G_0(\tau)$ -- the Manley-Raw relation for this case).
Therefore Eq. (\ref {eq11}) substantially simplifies:

\begin {eqnarray} \label {eq14}
\frac{\partial\theta}{\partial\tau}+\frac{G^2_0(\tau)}{K}
\frac{\partial\theta}{\partial z}=0.
\end {eqnarray}
The solution of Eq. (\ref {eq14}) can be written in the following
form:
\begin {eqnarray} \label {eq15}
\theta(\tau,z)=\theta_0(Z^{-1}(Z(\tau)-z),0),
\end {eqnarray}
where $Z(\tau)=K^{-1}\int_{-\infty}^{\tau}d\tau^{'}
G^2(0,\tau^{'})$, $Z^{-1}(\tau)$ is the inverse function of
$Z(\tau)$.

It is not difficult to show that in this case
$G=\sqrt{G_1^2(\tau,z)+G_2^2(\tau,z)}$ coincides with the
definition of dressed field pulses \cite{2,3}: $G_{-}=a_0 G_2 -
a_2 G_1$. Thus the pulses in our case can be identified as dressed
field pulses (only at $K_1=K_2$). It would be interesting to note
that other combination  $G_{+}=a_0 G_1+a_2 G_2\equiv 0$ (see also
\cite{2}). The concept of dressed field pulses cannot be applied
to the case of $K_1\neq K_2$, but in both cases the Manley-Raw
relation remains valid.

\subsection{The adiabaticity criterion}

The above results were obtained on the assumption that the
adiabaticity criterion (\ref {e1}) (or $|\dot{\theta}/G|\ll1$)
remains valid during propagation of pulses. However it is not
necessary the case. Therefore we investigate the adiabaticity
condition with the account of propagation. Differentiating (\ref
{eq13}) on $\tau$, we can write the following expression for the
adiabaticity criterion:

\begin {eqnarray} \label {eq16}
 \frac{\dot{\theta}(\tau,z)}{G(\tau,z)} &=&
  \frac{\partial\theta_0}{\partial\tau_0}
  \frac{G(\tau,z)}{G^2_0(\tau_0)}
  \frac{K(\theta(\tau,z))}{K(\theta_0(\tau_0))} \nonumber \\
  &\times& [1+\sin{(2\theta_0(\tau_0))}\frac{2(K_2-K_1)z}{G^2_0(\tau_0)}
  \frac{\partial{\theta_0}}{\partial{\tau_0}}]^{-1}
 \ll1
\end {eqnarray}
As follows from (\ref {eq16}), the adiabaticity condition is
destroyed ($\dot{\theta}/G \rightarrow \infty$) when the factor in
the square brackets tends to zero. Since
$\sin{(2\theta_0(\tau_0))}>0$ in the entire range of change of
$\theta$ ($0\leq\theta\leq\pi/4$), relation (\ref{eq16})  is not
fulfilled under the following conditions

\begin {eqnarray} \label {eq18}
\frac{2(K_1-K_2)z}{G^2_0(\tau_0)}\frac{\partial{\theta_0}}
{\partial{\tau_0}}\sin{(2\theta_0(\tau_0))}=1, \nonumber \\
(K_1-K_2)\frac{\partial{\theta_0}}{\partial{\tau_0}}>0,
\end {eqnarray}

Evidently, condition (\ref {eq18}) is not fulfilled at $K_1=K_2$,
and the adiabaticity criterion holds throughout the propagation
process in this case. It should be mentioned that numerical
analysis of Eq. (\ref{eq18}) reveals that, generally speaking,
there is a range of change of the $q=K_1/K_2$ parameter when the
first condition in (\ref{eq18}) is not fulfilled either: $q_{min}<
q < q_{max}$, $q_{min}<1$, $q_{max}>1$. The values $q_{max}$ and
$q_{min}$ depend on the ratio $a = T_2/T_1$ and the shape of
pulses. For example, the $q$ parameter for Gaussian pulses with
$T_2/T_1=3$ may vary within the limit $0.40<q<1.35$, as shown by
the numerical simulation.

In this case for the thick medium the adiabaticity criterion has
the form:
\begin {eqnarray} \label {eq19}
\frac{\partial\theta_0}{\partial\tau_0}\frac{G(\tau,z)}{G^2_0(\tau_0)}\ll1.
\end {eqnarray}
It can be readily shown that condition (\ref{eq19}) for Gaussian
pulses is fulfilled when $T_2/T_1>\sqrt{3}$ and $G_2^0T_1\gg1$
(compare with the case of the thin medium).

For Gaussian pulses ($T_2>T_1$) and the initial conditions
$a_0(-\infty)=1$, $a_2(-\infty)=0$, we have $\dot{\theta_0}>0$ for
$-\infty<\tau_0<0$.  With $K_1>K_2$ ($q>q_{max}$) the adiabaticity
criterion begins to break down at the leading edge: the mixing
angle front becomes steeper. Condition (\ref {eq16}) gets
destroyed at the trailing edge (where $\dot{\theta_0}<0$) when
$K_1<K_2$ ($q<q_{min}$). Let's introduce a critical length $z_c$ at
which the adiabaticity condition (\ref{eq16}) cannot be satisfied.
Using (\ref {eq18}), one can obtain the following simple
estimation for $z_c$ for the case $K_1 \neq K_2$:

\begin {eqnarray} \label {eq18a}
z_c \simeq \frac{G_0^2 T_1}{2|K_1-K_2|}.
\end {eqnarray}

Let us now consider the experimental parameters: $N = 10^{15}
cm^{-3}$; $1/\lambda_{1,2} = 10000, 20000 (cm^{-1})$;
$1/\gamma_{1,2} = 10, 50 ns$ are the relaxation times of levels
$|1\rangle$ and $|2\rangle$, respectively; $T_1 = 0.1 ns$;
$G_{1,2}^0T_1 = 20$; $a = T_2/T_1 = 3$; $fg_{10} = 0.1$ is the
oscillator strength of the probe transition. From (\ref{eq18a}) we
obtain $z_c = 10^5 z_0 \approx 2 cm$, where $z_0$ is the linear
absorption length. This estimation agrees with the results
presented in Figs. 5-7.

Using the characteristic equation (\ref{eq12}) it is easy to find
the area where the adiabaticity criterion breaks down.
Characteristic curves for various values of $q=K_1/K_2$ are shown
in Fig~4. The thickening of the characteristic curves means the
sharpening of the mixing angle front at a certain medium depth. At
the point of intersection (see Fig.~4~b,c), $\dot{\theta}
\rightarrow \infty$, condition (\ref{eq16}) collapses.  However
there are no such points in the case when $q_{min}<q<q_{max}$ (see
Fig.4a), i.e. here the adiabaticity criterion (\ref{eq16}) is
maintained during propagation of pulses.

The above discussion is illustrated by Fig.5 where the temporal
behavior of the mixing angle $\theta$ is presented for
$q=K_1/K_2=1,0.25,4$ and for different normalized propagation
lengths. On the Fig.5a-b the analytical results are presented
(formulae (\ref {eq13}) and (\ref {eq15})). Figure 5c shows the
numerical solution for $\theta$ at $K_1>K_2$.

One can see that the evolution of the $\theta$ parameter at
$K_1>K_2$ is different from that at $K_1<K_2$. At $K_1>K_2$ the
adiabaticity condition fails for all values of $z$, beginning from
the critical lengths $z_c$ defined by Eq. (\ref {eq18a}). Here,
the analytical theory does not apply at the very late stage of
nonlinear wave propagation. In the case of $K_1<K_2$ ,
nonadiabaticity develops at the trailing edge (the front becomes
steeper at a certain propagation length), but it does not go
deeper into the medium. A good agreement between the analytical
and numerical solutions for $\theta$ at $K_1\leq K_2$  is observed
over the entire propagation length. This leads us to conclude that
the interaction adiabaticity is fairly sensitive to the ratio of
the oscillator strengths of the transitions interacting with the
pulses.

\subsection{Discussion of results}

The solutions obtained have been used to analyze the temporal
dynamics and spatial behavior of the propagating EIT pulses and
the atomic coherence for various oscillator strength ratios.
Figure 6 illustrates the temporal and spatial evolution of
normalized Rabi frequencies of the both pulses
$g_{1,2}(\tau)=G_{1,2}/\sqrt{{G_1^0}^2+{G_2^0}^2}$ as they
propagate inside the optically thick medium. The temporal
evolution of pulses can be seen to depend on the ratio between the
transition oscillator strengths. In the case of $K_1\leq K_2$ both
pulses undergo reshaping as they propagate in the medium (Fig.6a
and b). The probe pulse is gradually depleted and the coupling gets
stronger. Note that the pulse shape at the initial stage of
propagation shows very little change along the length of the
medium, which may exceed the linear absorption length. Complete
reemitting of the probe pulse into the coupling one during
propagation is possible. Using the Eq.  (\ref {eq12}) one can
obtain the following expression for the maximal distance $z_m$ on
which the probe pulse propagates into the medium:

\begin{eqnarray} \label {eq20}
 z_m=\frac{1}{K^2(\theta_0(-\infty))}
 \int_{-\infty}^{\infty}A(\tau)d\tau.
\end{eqnarray}
For the Gaussian pulses we  have

\begin{eqnarray} \label {eq21}
z_m = \sqrt{\frac{\pi}{2ln{2}}}\,(G_2^0)^2\; T_1
 \frac{K_2(G_1^0/G_2^0)^2+K_1a}{K_1^2}.
\end{eqnarray}
Here $a=T_2/T_1$.

For the same parameters as in the previous sub-section, we have
 $z_m = 4.8 \cdot 10^5 z_0 \approx 9 cm$ for $K_1/K_2 = 1$ and
 $z_m = 8.4 \cdot 10^5 z_0 \approx 16 cm$ for $K_1/K_2 = 1/4$.
These values agree with the results shown in Figs. 5-7.

An interesting feature of the spatial distribution of the probe
pulse is illustrated in Fig. 6a,b: in some areas of the medium, the
field in the tail of the pulse  is different from zero. This is
believed to be the result of the spatial compression of the probe
caused by the slowing down of the group velocity of the probe pulse
\cite{Lukin}.

In the case of $K_1>K_2$, the adiabaticity condition is maintained
over the $z_c$ range, which can also be much longer than the
length of the linear absorption. The leading edge of the probe
pulse undergoes gradual depletion and the pulse amplitudes display
only small changes in that range. Outside that range, the pulse
splits into several peaks. This occurs due to the nonadiabaticity
of interaction (see Fig. 6c where numerical simulations are
presented for $g_{1,2}(\tau)$. Here the first three curves
correspond to the adiabatic interaction).

The pulse evolution, described above, is determined by spatial and
temporal behaviour of the atomic Raman coherence $\rho_{20}$ shown
in Fig.7 (see also \cite{7a}). Figures 7a,b reveal an unusual
spatial and temporal behavior of the atomic coherence, which we
interpret as the slowing down, stopping and localization of the
atomic coherence in the medium. The probe pulse is transferred
into and stored in the collective atomic excitation under the
control of the coupling. One can say that a phase grating is
created in the atomic medium. The phase grating is preserved
throughout the entire period of relaxation of the atomic
coherence. During that period of time, the information stored in
the atomic excitations can be transferred back to the radiation
using another coupling pulse of the same or of a different
frequency.

We find these effects to be similar to the ones predicted and
demonstrated in \cite{Phillips,Liu,Lukin}, but for some
differences as indicated below. In our case, both the coupling
pulse and the probe pulse are strong. Unlike \cite{Lukin}, we used
the boundary conditions for the pulse envelopes yielding  the time
distribution of pulses at the medium boundary, $z=0$. We believe,
this condition is more natural than the one used in \cite{Lukin}
where the authors use the probe pulse distribution in the medium
at fixed time as the initial condition. Also they do not take into
consideration evolution of the coupling pulse. The effect takes
place in the case $K_1\leq K_2$ and is not observed in the case of
$K_1>K_2$.

\section{CONCLUSION}

The propagation of two short overlapping pulses with durations
$T_2>T_1$ in optically thick three-level media under CPT
conditions has been studied for the general case of unequal
transition oscillator strengths. An analytical solution has been
obtained for the set of reduced wave equations under the adiabatic
following condition. Also it has been shown how spatial evolution
of pulses depends on the oscillator strength ratio.

The condition of adiabaticity provided at the medium entrance
preserves for any value of propagation lengths if $K_1\leq K_2$
($q<q_{max}$) and breaks down at $K_1>K_2$ ($q>q_{max}$). In the
range $q<q_{max}$, the probe pulse is completely depleted and
reemitted into the coupling pulse during propagation.  This is not
possible in the case of $q>q_{max}$. It has been established that
to provide for the adiabaticity condition in an optically thin
medium, the restriction $T_2/T_1> \sqrt{2}$ has to be ensured,
whereas in a thick medium $T_2/T_1> \sqrt{3}$.

We also have studied the spatial behavior of the atomic coherence
$\rho_{20}$, which plays a significant role, for example, in
nonlinear mixing processes. It has been found that a strong
coherence can be maintained over a length equal to several
hundreds of thousand of one-photon absorption lengths during
propagation. The effect of localization of the atomic coherence is
demonstrated.


\newpage
\centerline{FIGURES}

FIG. 1. (a) The three-level Lambda-type system coupled by two
resonant pulses with Rabi frequencies $G_1$ and $G_2$. (b) The
shapes of probe $G_1$ and coupling $G_2$ pulses at the medium
entrance.

FIG. 2. The time evolution of (a) level populations
$|a_{0,2}(\tau)|^2$ and the atomic Raman coherence
$|\rho_{20}(\tau)|$, (b) the mixing angle $\theta(\tau)$ in an
optically thin medium for the Gaussian pulses. $G_1^0 T_1=G_2^0
T_1=20$, (a) $T_2/T_1=3$; (b) $T_2/T_1=\sqrt{2}$ (A), $\sqrt{3}$
(B), $3$ (C), $10$ (D).

FIG. 3. The vector model of adiabatic interaction of two short
pulses with a three-level  Lambda-type system.

FIG. 4. The characteristic curves for the Eq.(\ref{eq11}): (a)
$K_1/K_2=1$, (b) $K_1/K_2=0.25$, (c) $K_1/K_2=4$.

FIG. 5. The time evolution of the parameter $\theta$ for different
relationships between $K_1$ and $K_2$ at different propagation
lengths.  (a) $K_1=K_2$, (b) $K_1/K_2=0.25$, (c) $K_1/K_2=4$.
$G_{1,2}^0 T_1=20$, $T_2/T_1=3$. Here and in all the other figures
the time $\tau$ is measured in the units of the pulse duration
$T_1$, and the propagation length $z$ of pulses in the medium is
measured in the units of the length of linear absorption of the
probe radiation determined in accordance with the Beer's low. In
(c) the numerical solution for the case $K_1/K_2=4$ is presented.

FIG. 6. The time evolution of the normalized Rabi frequencies
$g_{1,2}=G_{1,2}/\sqrt{{G_1^0}^2+{G_2^0}^2}$ of the probe and
coupling pulses for different relationships between $K_1$ and
$K_2$ at different propagation lengths.  (a) $K_1=K_2$, (b)
$K_1/K_2=0.25$, (c) $K_1/K_2=4$. $g_{1,2}^0=20$, $T_2/T_1=3$.

FIG. 7. The time evolution of the atomic Raman coherence for
different relationships between $K_1$ and $K_2$ at different
propagation lengths.  (a) $K_1=K_2$, (b) $K_1/K_2=0.25$, (c)
$K_1/K_2=4$. $G_{1,2}^0T_1 =20$, $T_2/T_1=3$.

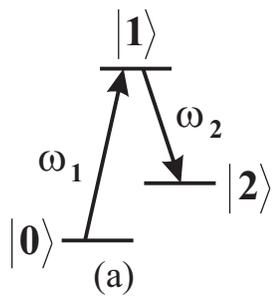
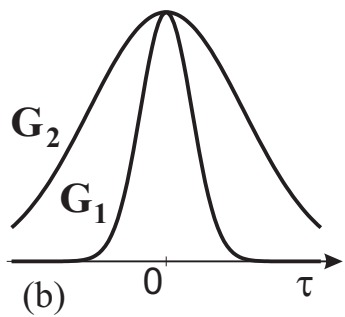
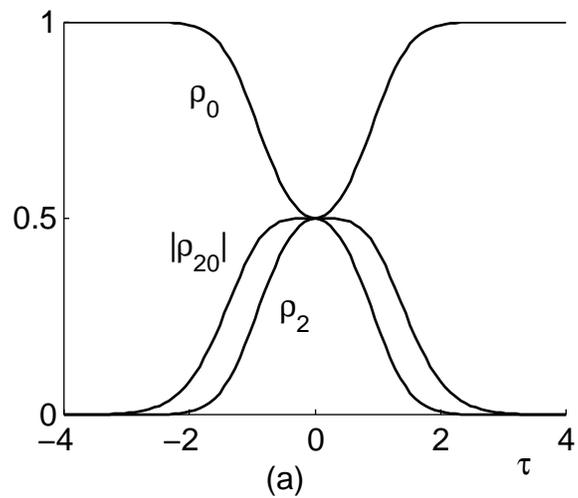
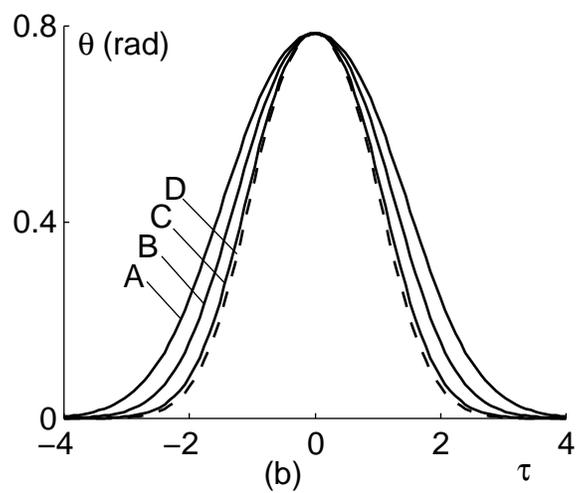
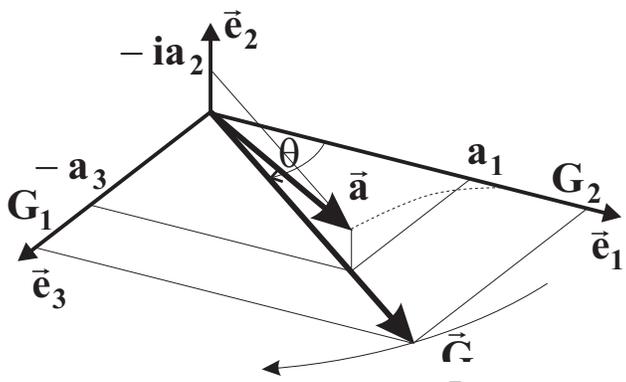
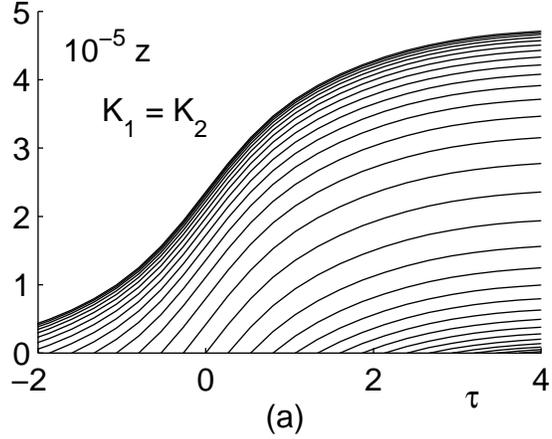
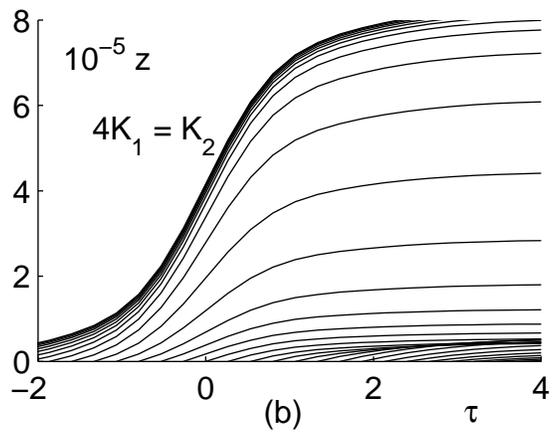
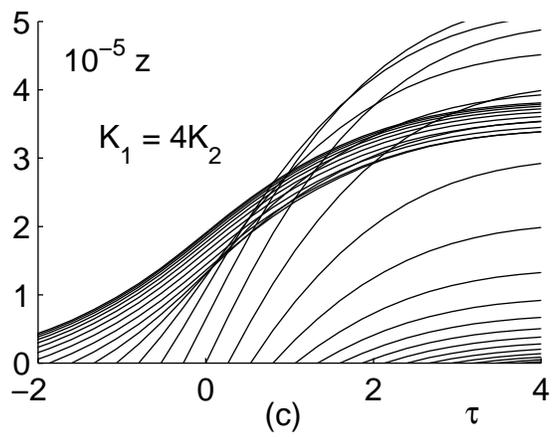
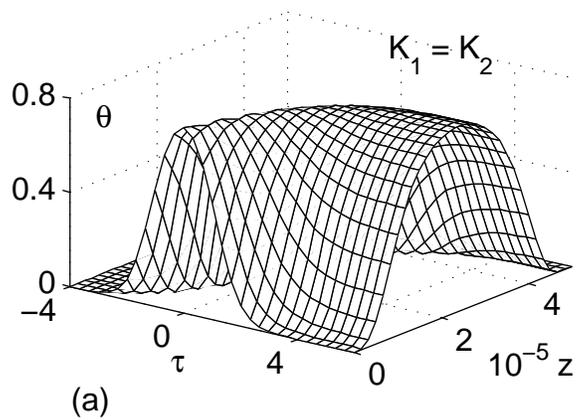
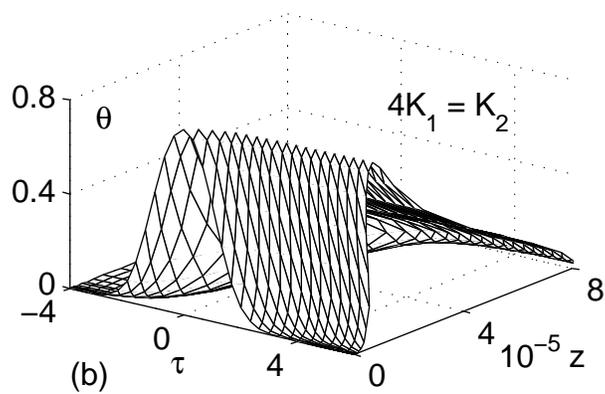
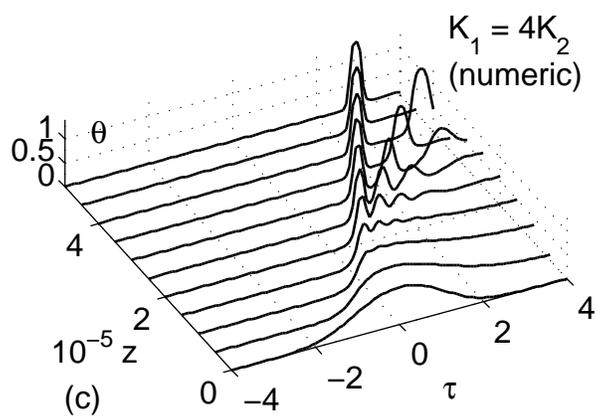
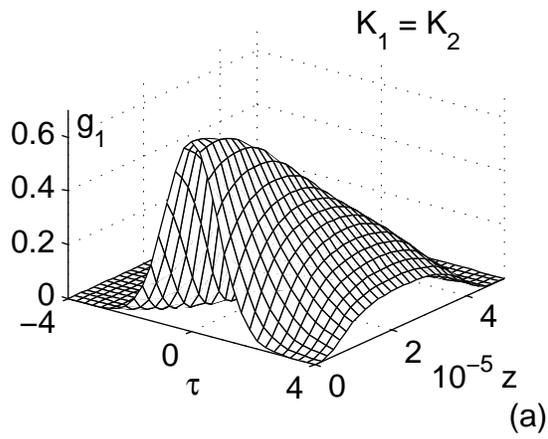
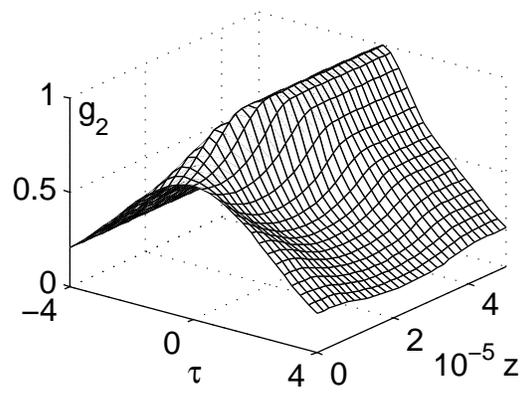
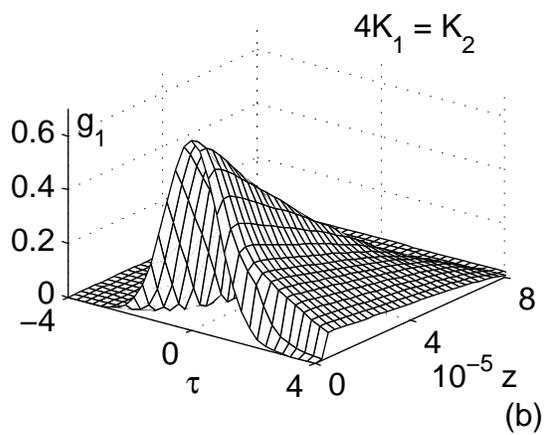
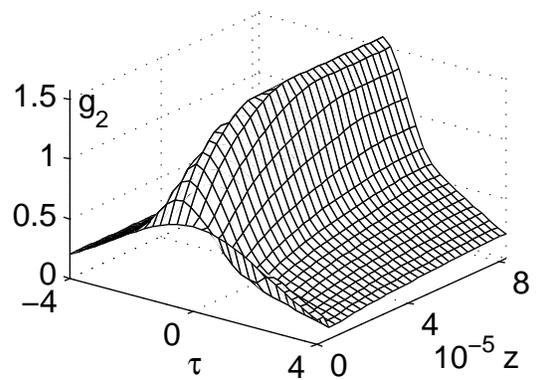
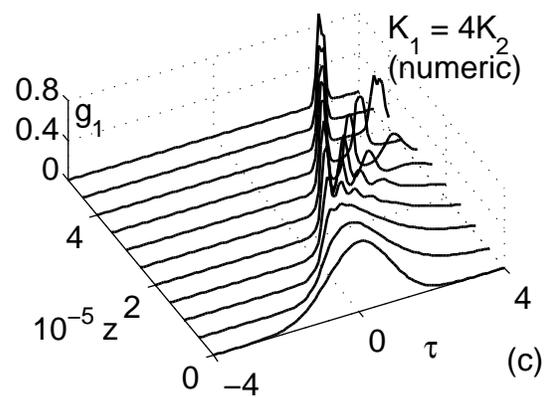
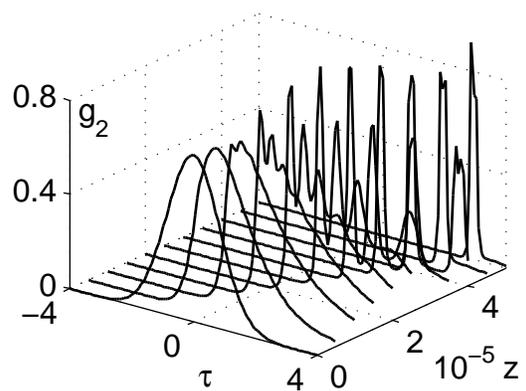
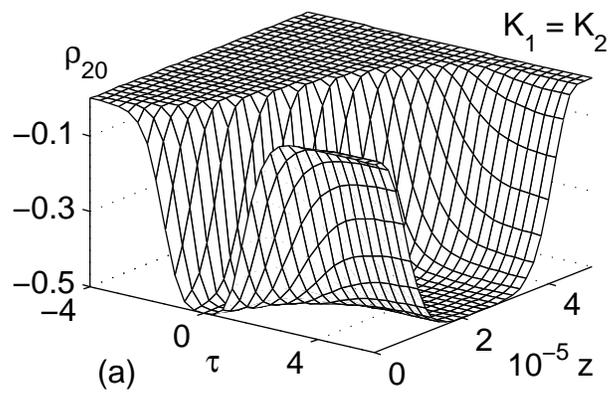
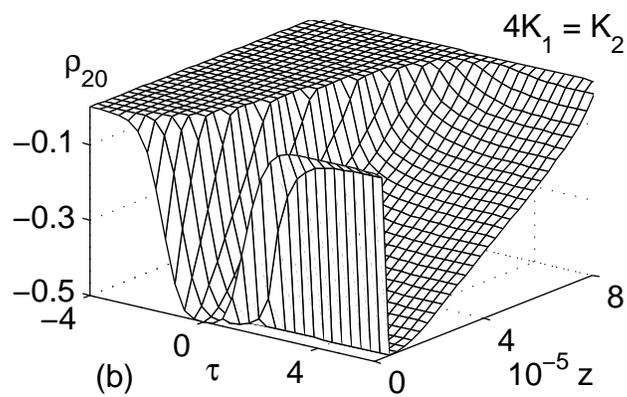
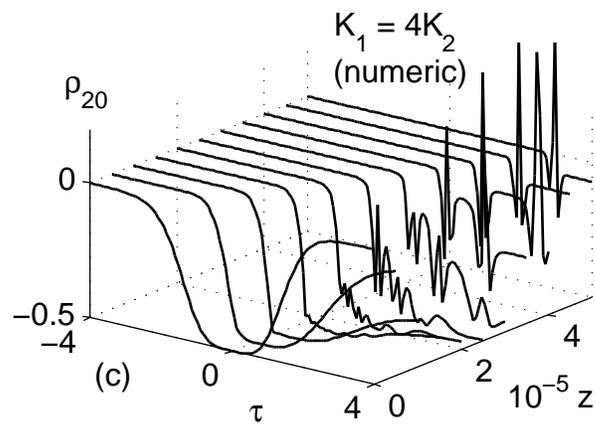
\end{document}